\documentclass[preprint,12pt]{elsarticle}

\usepackage{graphicx}

\usepackage{gensymb}
 
\usepackage{amssymb}

\usepackage{textcomp} 
\usepackage{color,soul}
\usepackage{txfonts}

\journal{arxiv}

\begin{document}\sloppy

\begin{frontmatter}


\cortext[cor1]{thomas.day.goodacre@cern.ch}

\title{Identification of autoionizing states of atomic chromium for resonance photo-ionization at the ISOLDE-RILIS}


\author[1,2]{T.~Day~Goodacre*}
\author[1,4]{K.~Chrysalidis}
\author[3]{D.~Fedorov}
\author[1]{V.~N.~Fedosseev}
\author[1]{B.~A.~Marsh}
\author[3]{P.~Molkanov}
\author[1,4,5]{R.~E.~Rossel}
\author[1]{S.~Rothe}
\author[1]{C.~Seiffert}

\address[1]{\scriptsize{CERN, CH-1211 Geneva 23, Switzerland}}
\address[2]{\scriptsize{School of Physics and Astronomy, The University of Manchester, Manchester, M13 9PL, United Kingdom}}
\address[3]{\scriptsize{Petersburg Nuclear Physics Institute, 188350 Gatchina, Russia}}
\address[4]{\scriptsize{ Institut f{\"u}r Physik, Johannes Gutenberg Universit{\"a}t, D-55099 Mainz, Germany}}
\address[5]{\scriptsize{Faculty of Design, Computer Science and Media, Hochschule RheinMain, Wiesbaden, Germany}}

\begin{abstract}
The resonance ionization laser ion source (RILIS) is the principal ion source of the ISOLDE radioactive beam facility based at CERN. Using the method of in-source resonance ionization spectroscopy, an optimal three-step, three-resonance photo-ionization scheme has been developed for chromium. The scheme uses an ionizing transition to one of the 14 newly observed autoionizing states.  This work increases the range of ISOLDE-RILIS ionized beams to 32 chemical elements. Details of the spectroscopic studies are described and the new ionization scheme is summarized. A link to the complete version of this document will be added here following publication:\\

\end{abstract}


\end{frontmatter}


\section{Introduction}
\label{Introduction}

\subsection{The ISOLDE-RILIS}
ISOLDE is an isotope separator on-line (ISOL) type, radioactive ion beam facility connected to the CERN accelerator complex~\cite{Kugler2000a}. Radionuclides are produced with a pulsed 1.4 GeV proton beam of up to 2~\textmu A average current that is impacted upon a "thick target". Reaction products, created via fragmentation, spallation and fission, are stopped and thermalized within the target material. The target is heated to enable sufficiently volatile elements to diffuse through the target material and then effuse, via a transfer line, to an ion source typically for 1+ ionization. 

The resonance ionization laser ion source (RILIS) is the principal ion source of the ISOLDE facility, used for 75$\%$ of the ISOLDE physics experiments in 2015. The hot cavity RILIS operation at ISOLDE is based on the following three principles:
\begin{enumerate}
	\item The distribution of the atomic energy levels is unique to a particular element.
	\item Electric dipole transitions between electronic states of the atom can be driven by an external photon field, when the photon energy corresponds to the transition energy. 
	\item Ions in the hot cavity environment (typically a 3~mm internal diameter metal tube heated up to 2400~K) are transversally confined by the cavity walls, which are positively charged as a result of thermally induced electron emission, enhancing ion survival rates.   
\end{enumerate}

\noindent The RILIS targets a progressive series of resonant atomic excitations, using multiple tunable lasers, before a final ionizing step: either a resonant transition to an autoionizing state or non-resonant ionization to the continuum~\cite{Letokhov1987}. Ionization via an autoionizing resonance is preferred as the photo-absorption cross-section for resonant transitions can  be orders of magnitude higher than for non-resonant ionization, often making saturation of the transition feasible with the typically available laser power~\cite{mishin1993}.

The development of a RILIS scheme for chromium was initiated following interest from the ISOLDE solid state physics group and the ISOLTRAP collaboration for experiments involving neutron deficient and neutron rich isotopes respectively~\cite{Johnston2015, Kreim2011}. Previous chromium production at ISOLDE used arc discharge ion sources, most recently a VD5 "hot plasma" variant of the ISOLDE Versatile Arc Discharge Ion Source (VADIS), coupled to a ZrO$_{2}$ or a Y$_{2}$O$_{3}$ target~\cite{Stora2013}. The "universal" ionization efficiency of the VD5 VADIS limits the achievable beam purity for some isotopes due to isobaric contamination, particularly if a UC$_{X}$ target is used~\cite{Kirchner1981}. A viable RILIS scheme therefore enhances the purity of chromium ion beams. In addition to offering the other advantages of element selective ionization with the RILIS, such as the ability to remove the laser ionized chromium component of the ion beam by blocking one of the resonant transitions, enabling  measurements to be made both with the lasers blocked and unblocked for signal identification.

\section{Ionization of chromium atoms in a hot cavity}

Chromium has a first ionization potential of 6.77~eV~\cite{Kramida2014}. Consequently, with sufficient heating of the sample of stable chromium, a surface ion signal measurable on an ISOLDE Faraday cup ($>$100~fA) is expected from a tungsten hot cavity ion source heated to 2400~K.  

\section{Experimental set-up}

The RILIS scheme development for chromium took place at ISOLDE using the RILIS laser system and a tungsten hot cavity ion source mounted on the general purpose separator (GPS) frontend. The ISOLDE-RILIS is primarily made up of broadly tunable lasers to maximize the spectral range and flexibility of the of the system, with a combination of three dye lasers and three Titanium:Sapphire (Ti:Sa) lasers. An independent 40~W average power, TEM$_{00}$, frequency doubled Nd:YVO$_{4}$ (532~nm) Lumera Blaze laser, is available for ionization schemes using a non-resonant final ionizing step~\cite{Marsh2013}. The lasers are pulsed at a repetition rate of 10~kHz, approximately the frequency required to ensure atoms effusing through the ion source are illuminated by a minimum of one set of laser pulses. The tunable lasers are typically pumped by frequency doubled Nd:YAG (532~nm) lasers; however, the dye lasers can also be pumped with the third harmonic of the Nd:Yag (355~nm) to extend their possible tuning range. Through the use of non-linear optics for 2nd, 3rd and 4th harmonic generation, an unbroken spectral coverage between 210~nm and 950~nm is achieved. The wavelengths of the RILIS lasers are measured using a HighFinesse/\AA ngstrom WS7 wavelength meter. Two dye lasers, one frequency doubled Ti:Sa laser and the 532~nm Blaze laser, highlighted in the schematic of the RILIS laser system presented in Figure~1, were used during the scheme development of chromium.  

\begin{figure*}[htb] 
\vspace{-3mm}
\includegraphics[width=\textwidth]{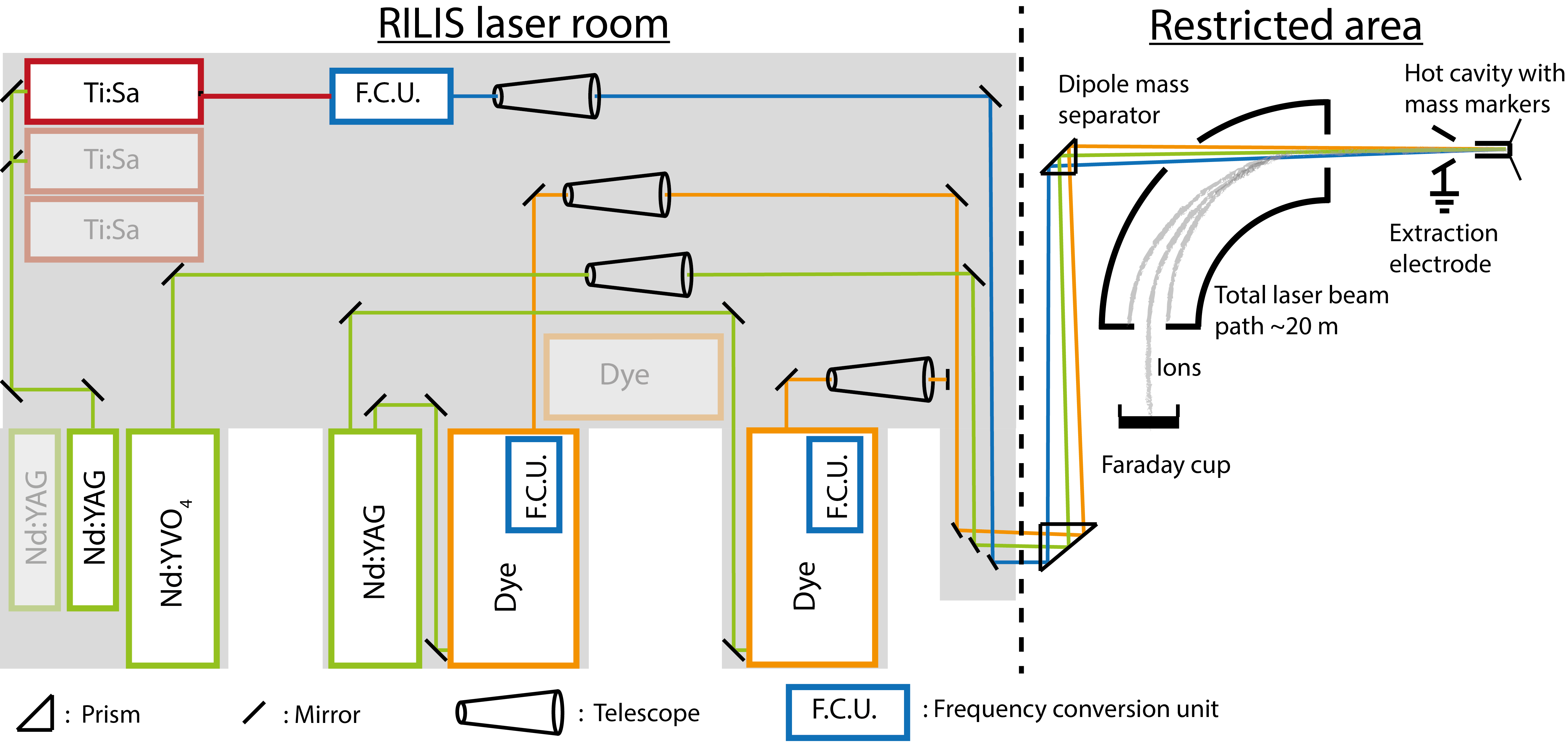}
\caption [Schematic of the RILIS layout and the experimental set-up for the development of a RILIS scheme for chromium]{\footnotesize Schematic of the RILIS layout and the experimental set-up for the development of a RILIS scheme for chromium} 
\label{fig:LT}
\end{figure*}

For RILIS scheme development, capillary mass markers loaded with stable isotopes of the element of interest are connected to the rear of a hot cavity, providing a continuous supply of atoms within the ion source as they are heated and evaporated. As depicted in Figure~\ref{fig:LT}, the RILIS laser beams were directed through the GPS dipole magnet, to converge within the hot cavity ion source. The ion source assembly was maintained at a voltage of 30~kV and a grounded extraction electrode was positioned $\sim$ 60~mm from the exit of the ion source. Ions created in the ion source drifted towards the cavity exit under the influence of a longitudinal voltage drop resulting from the resistive heating of the hot cavity. At the exit of the hot cavity, the ions were extracted by the penetrating field of the extraction electrode and accelerated to form a 30~keV ion beam. The ions passed through the ISOLDE GPS dipole magnet for mass selection before being directed to a Faraday cup for measurement of the ion current. 

The 357.9~nm light for the first step transition was produced by frequency doubling the output of a Ti:Sa laser and the 697.8~nm light for the second step transition was produced by a dye laser containing Pyridine 1 dye dissolved in ethanol. A second dye laser, containing Rhodamine 6G dye dissolved in ethanol, was used applied as a third excitation step to above the first ionization potential and scanned to search for autoionizing states. The $^{52}$Cr (stable) ion current was measured whilst recording the scanning laser wavelength using the RILIS data acquisition system~\cite{Rossel2013a}. Since the purpose of the spectroscopic study was the identification of an optimal RILIS scheme for chromium beam production at ISOLDE, typical RILIS operating conditions were maintained (laser linewidths and power levels were not optimized for spectral resolution). The RILIS laser parameters are described in detail by Fedosseev et al. and Rothe et al.~\cite{Fedosseev2012c, Rothe}. 

\section{Results and discussion}

The procedure used for laser resonance ionization scheme development for the ISOLDE-RILIS was reviewed by Fedosseev et al.~\cite{Fedosseev}. According to Boltzmann's equation, the 3d$^{5}$($^{6}$S)4s a$^{7}$S$_{3}$ atomic ground state of chromium is 97~$\%$ populated at 2400~K. Thus, all of the RILIS schemes considered here originate from the atomic ground state. Extensive atomic line data is available for chromium, including numerous autoionizing states~\cite{Smith, Kramida2014, Saloman2012}. Following a review of these atomic lines and energy levels, suitable first and second steps for a RILIS scheme were identified, details of the transitions are presented in Table~\ref{tab:1}.\\

\begin{table*}[htb]
  \vspace{-5mm}
	\centering
  \caption{The first and second resonant transitions used in the chromium scheme development. Spectroscopic information is taken from NIST database~\cite{Kramida2014}}
	\begin{tabular}{ c c c c c }
	\hline
	Transition & Upper state config., term, J & Wavenumber &Air wavelength\\
	(cm$^{-1}$) & & (cm$^{-1}$) & (nm) & \\  
	\hline
0 - 27935 & 3d$^{4}$($^{5}$D)4s4p$^{3}$P$^{\circ}$, y$^{7}$P$^{\circ}$, 4 &	27935 & 357.9\\
27935 - 42261   & 3d$^{5}$($^{6}$S)4d, e$^{7}$D, 5 &	14325 & 697.8\\
  \end{tabular}
  \label{tab:1}
\end{table*}

The upper level of the second transition lies at an energy of 42261 cm-1, $\sim$12314~cm$^{-1}$ ($\sim$812~nm) below the ionization threshold~\cite{Kramida2014}. When considering the suitability of a transition, information concerning the photoabsorption cross-section of the candidate transition was compared to similar data for transitions used in existing RILIS schemes, for which a saturation measurement has been performed.  This provides a reliable estimate of the feasibility of saturating a proposed transition with the laser power that is typically available from the RILIS lasers. Where available, an excitation scheme consisting of states with progressively increasing statistical weights is preferred, as this increases the maximum achievable ionization efficiency in the RILIS ionization geometry.

Initially a three step, two resonance scheme, with a non-resonant final step at 532~nm was demonstrated ($\{\lambda_{1}\left|\lambda_{2}\right|\lambda_{3}\} = \{357.9~$nm$\vert 697.8~$nm$\vert532~$nm$^{NR}\}$), verifying the suitability of the transitions found in literature. The non-resonant final step was then replaced with the tunable dye laser containing Rhodamine 6G dye dissolved in ethanol to excite from the 3d$^{5}$($^{6}$S)4d e$^{7}$D$_{5}$ level at 42261 cm$^{-1}$ and scan above the ionization potential. 14~new autoionizing states were observed in the scan between 17204~cm$^{-1}$ and 17854~cm$^{-1}$.

\begin{table*}
  \vspace{-5mm}
	\centering
 \caption{An autoionizing states of chromium, accessed by a single photon transition from the 3d$^{5}$($^{6}$S)4d e$^{7}$D$_{5}$ level at 42261 cm$^{-1}$.} 
 \begin{tabular}{ c c c c c }
	\hline
	 Energy & Transition & Air wavelength \\
	(cm$^{-1}$) & (cm$^{-1}$) & (nm) \\  
	\hline
59523 & 17262 & 579.1	\\
  \end{tabular}
 
  \label{tab:crais}
\end{table*}

\begin{figure*}[htb] 
\vspace{-3mm}
\includegraphics[width=\textwidth]{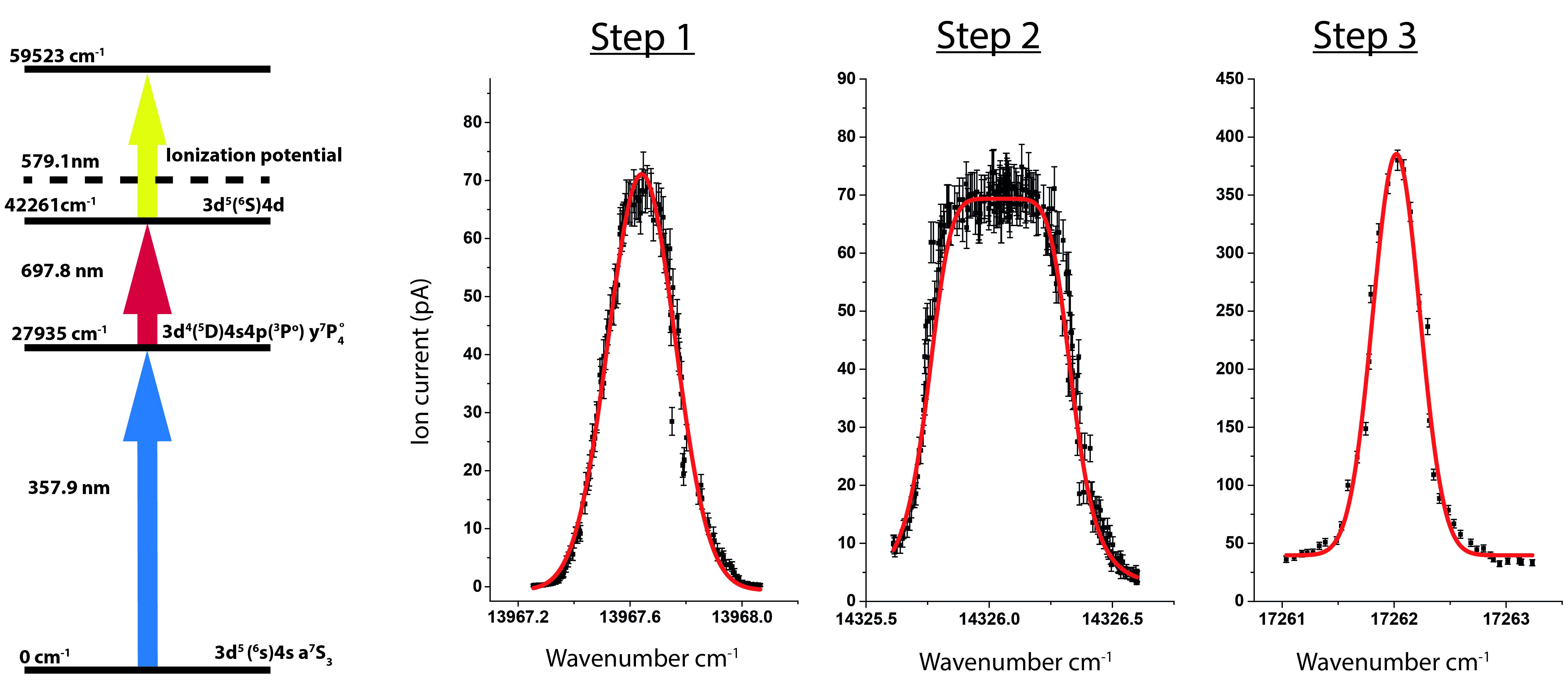}
\caption [The optimal RILIS ionization scheme for chromium and scans of the three resonant transitions.]{\footnotesize The optimal RILIS ionization scheme for chromium and scans of the three resonant transitions.} 
\label{fig:crsns}
\end{figure*}

The optimal scheme, corresponding to the highest observed $^{52}$Cr ion current, is presented in Figure~\ref{fig:crsns}, along with laser scans of the three resonances.   

Time constraints limited the investigation of saturation levels to only the $\{357.9~$nm$\vert 697.8~$nm$\vert 579.1~$nm$^{AI}\}$ scheme. The saturation of the first two resonant transitions was verified and a saturation measurement of the third resonant transition was made by measuring the $^{52}$Cr ion current while varying the third step laser power, the transition was not saturated with an estimated 1.3 W of laser light transmitted to the ion source. A redistribution of the pump laser power for the dye lasers could enable up to 7~W of laser power at 579~nm to be delivered to the ion source, increasing the total efficiency of the scheme.

Using the scheme depicted in Figure~\ref{fig:crsns}, a laser to surface ionization ratio of 2200:1 was recorded with the hot cavity heated to 2400~K. The ion signal when applying the $\{357.9~$nm$\vert 697.8~$nm$\vert 579.1~$nm$^{AI}\}$ scheme, was a factor of two greater than the ion signal observed when using the $\{357.9~$nm$\vert 697.8~$nm$\vert532~$nm$^{NR}\}$ scheme with non-resonant ionization despite an estimated 24~W of 532~nm laser light transmitted to the ion source. 

Chromium extraction from ISOLDE UCx targets has been verified, using surface ionization, as far in to the neutron rich region as $^{59}$Cr~\cite{Althubiti2015}. This demonstrates the possibility of extracting chromium isotopes with half lives of the order of 100~ms. The validity of the new RILIS scheme has been demonstrated on-line at ISOLDE with the measurement of RILIS ionized, radiogenically produced $^{52}$Cr. A factor of 800 enhancement of the ion signal was observed, from an ion beam composed of multiple elements with the line heated to 2400~K.

\section{Conclusion}

14 new autoionizing states of chromium have been observed, during off-line testing at the ISOLDE-RILIS. From this work, an optimal $\{357.9~$nm$\vert 697.8~$nm$\vert 579.1~$nm$^{AI}\}$  RILIS scheme was identified. The scheme was verified on-line at ISOLDE, using radiogenically produced  chromium from a UC$_{X}$ target. As a result, RILIS ionized chromium beams are now available for experiments at ISOLDE.

\section{Acknowledgements} 
This project has received funding through the European Union's Seventh Framework Programme for Research and Technological Development under Grant Agreements 262010 (ENSAR), 267194 (COFUND), and 289191 (LA$^{3}$NET).



\bibliographystyle{unsrt}
\bibliography{cr}







\end{document}